\documentclass[12pt]{article}
\usepackage{mrnd}
\usepackage[dvips]{graphicx}
\usepackage{amsbsy}

\def\Journal#1#2#3#4{{#1} {\bf #2}, #3 (#4)}

\def\NPA{{\em Nucl. Phys.} A}
\def\NPB{{\em Nucl. Phys.} B}
\def\PLB{{\em Phys. Lett.}  B}

\def\PRL{\em Phys. Rev. Lett.}
\def\PRD{{\em Phys. Rev.} D}
\def\PRC{{\em Phys. Rev.} C}

\def\EPJA{{\em Eur. Phys. J.} A}

\def\be{\begin{equation}}
\def\ee{\end{equation}}
\def\bea{\begin{eqnarray}}
\def\eea{\end{eqnarray}}

\begin{document}
\vspace*{4cm}
\title{HIGHER TWISTS AND NUCLEAR EFFECTS}

\author{ S.A. KULAGIN }

\address{Institute for Nuclear Research of the Russian Academy of Sciences\\
Moscow 117312, Russia}

\maketitle\abstracts{
We discuss the impact of nuclear effects on the higher twist terms
using a particular example of the structure function $xF_3$ as extracted
from a QCD fit to neutrino deep inelastic data on a heavy nucleus target. }

The operator product expansion is a common theoretical framework in analyses 
of deep inelastic scattering (DIS) in QCD. The operators can be ordered according
to their twist yielding the series in $Q^{-2}$ for physical observables. 
For example, for the structure function $F_3$, this reads
\begin{eqnarray}
\label{t-ex}
xF_3(x,Q^2) = xF_3^\mathrm{LT}(x,Q^2)+\frac{H_3(x,Q^2)}{Q^2}+\ldots
\end{eqnarray}
The first term in this expansion (the leading twist, LT) dominates at sufficiently large
momentum transfer $Q^2$ and invariant mass $W^2=M^2+Q^2(1-x)/x$. The LT structure functions
are constructed in terms of 
parton distribution functions (PDFs), which are universal for charged lepton and neutrino
scattering and have clear probabilistic interpretation.
An accurate knowledge of these
plays a key role in the extraction of possible contributions of new physics at 
new collider energies, non-accelerator physics (cosmic neutrinos) and, 
as observed more recently, in the interpretation of forthcoming
high precision experiments on neutrino oscillation.

PDFs are not directly observable and extracted from data in global QCD fits
(see, e.g., \cite{a}). 
The higher twist (HT) terms are presently poorly known and 
currently is a subject of both theoretical
and phenomenological studies \cite{mv,kps,a,akl,ks}. A better understanding of HT
terms, in particular their role in describing low $Q^2$ and high $x$ DIS data,
is important and provides valuable information on quark--gluon correlations
inside the nucleon.

Phenomenological studies of PDF and HT terms are affected by a number of 
uncertainties. These include the effects due to 
higher order perturbative terms in renormalization group analysis, 
threshold resummation effects, 
target mass corrections,
the sensitivity of QCD fits to the choice of parameterization of PDFs and HT, etc.
If nuclear data are involved into an analysis, nuclear effects will also have impact on
the extracted PDF and HT terms.

In the present contribution we discuss the size of nuclear effects on a particular
example of the HT term in the structure function $F_3$ as extracted from
neutrino data of CCFR collaboration on iron target \cite{ccfr} 
(for more detailed discussion see \cite{ks}).
In our fit, we separate the 
nucleon structure function  $F^N_3(x,Q^2)$ into 
a sum of the LT and the HT terms, (\ref{t-ex}), and
parametrize $xF_{3}^\mathrm{LT}$ at some  scale $Q^2=Q_0^2$
in terms of a simple function,
\begin{eqnarray}
xF_3^\mathrm{LT}(x,Q^2_0) =a_1x^{a_2}(1-x)^{a_3}(1+a_4 x).
\label{xf30}
\end{eqnarray}
Then we apply the renormalization group equation in order to calculate evolution of
$xF_{3}^\mathrm{LT}$ with $Q^2$. We solve the renormalization group equation in the
leading (LO), next-to-leading (NLO) and next-to-next-to-leading (NNLO) logarithm
approximations of QCD. In doing so we expand the leading twist
structure function $xF_3^{LT}$ in terms of its
Mellin moments within the framework of the Jacobi polynomial method and then apply
the evolution equations to the moments 
(for more detail see \cite{kps} and references therein).

We fit 116 data points on the structure function $F_3$ from the CCFR experiment \cite{ccfr}
in the kinematical range of $0.0075\le x\le 0.75$
and $Q^2$ between 1.3 and 200\,GeV$^2$. 
The fit parameters are $a_2$, $a_3$, and $a_4$ of
Eq.(\ref{xf30}) at the scale $Q^2_0$, the values of the function $H_3$ at the center of
each $x_i$-bin of the CCFR data set, 
as well as the QCD scale parameter  $\Lambda_{\overline{MS}}$.
The parameter $a_1$ was fixed by normalizing (\ref{xf30}) to the
Gross-Llewellyn-Smith sum rule, which was calculated in QCD to the second order
 in $\alpha_S$ \cite{GL},
$S_{\rm GLS}= 3(1-\alpha_S/\pi-3.25(\alpha_S/\pi)^2)$.

It should be noticed that in general the higher twist terms can be of two kinds:
those which have the kinematical
nature, e.g. the terms due to finite target mass, and those which arise due to
higher twist operators and reflect the quark-gluon interaction effects in the
target.
In order to ensure that the HT term in (\ref{t-ex})
describes effects due to quark-gluon
interaction in the target
we explicitly take into account target mass correction by
substituting
the Mellin moments by the Nachtmann moments
in the Jacobi polynomial expansion
of $xF_{3}^\mathrm{LT}$.

In order to apply corrections for nuclear effects in our analysis we first calculate
the ``EMC ratio'' for the iron target, 
$R_3(x,Q^2)=\frac1{56}F_3^\mathrm{Fe}(x,Q^2)/ F_3^N(x,Q^2)$,
with $F_3^\mathrm{Fe}$ the structure function of the $^{56}\mathrm{Fe}_{26}$ nucleus 
and $F_3^N=\frac12(F_3^p+F_3^n)$ the structure function of an isolated isoscalar nucleon.
Then we extract ``data" on the structure function of the isoscalar nucleon
from the CCFR data as $F_3^N(x,Q^2)=F^{\rm CCFR}_3(x,Q^2)/R_3(x,Q^2)$.
In calculating the ratio $R_3$ we observe that the
bulk of neutrino data with $Q^2> 1$\,GeV$^2$ is located in the region of $x>0.1$.
For this kinematical regime it is usually assumed that nuclear DIS of leptons
from nuclear targets can be viewed as incoherent scattering from bound nucleons.
Major nuclear effects found in this region are due to
nuclear binding and Fermi motion \cite{binding}. The relation between a heavy nucleus
structure function and the proton and neutron structure functions can be written as
follows
\begin{eqnarray}
\label{xf3A}
xF_3^A = \left\langle
\left(1+\frac{p_z}{\gamma M}\right)
\left(x'F_3^N +
\frac{N{-}Z}{2A}\left(x'F_3^n-x'F_3^p\right)
\right)
\right\rangle,
\end{eqnarray}
where 
 $x'=Q^2/2p\cdot q$ is the Bjorken variable of the bound
nucleon with the four-momentum $p$, $q$ is four momentum transfer
 and $\gamma=|\boldsymbol{q}|/q_0$ is the `velocity'
of the virtual boson in the target rest frame. The averaging is done with respect to the
nuclear spectral function and the last term in Eq.(\ref{xf3A}) takes into account that 
the neutron, $N$, and the proton, $Z$, numbers are unequal in heavy nuclei.
Note also that bound proton and neutron are off-mass-shell and their structure
functions depend on the nucleon off-shellness $p^2$ as an additional variable
that causes an additional correction \cite{off-shell}. The nuclear spectral function 
and other details of the present approach are discussed in \cite{ks}.

\begin{figure}
\begin{center}
\vskip -2.5cm
 \includegraphics[width=\textwidth]{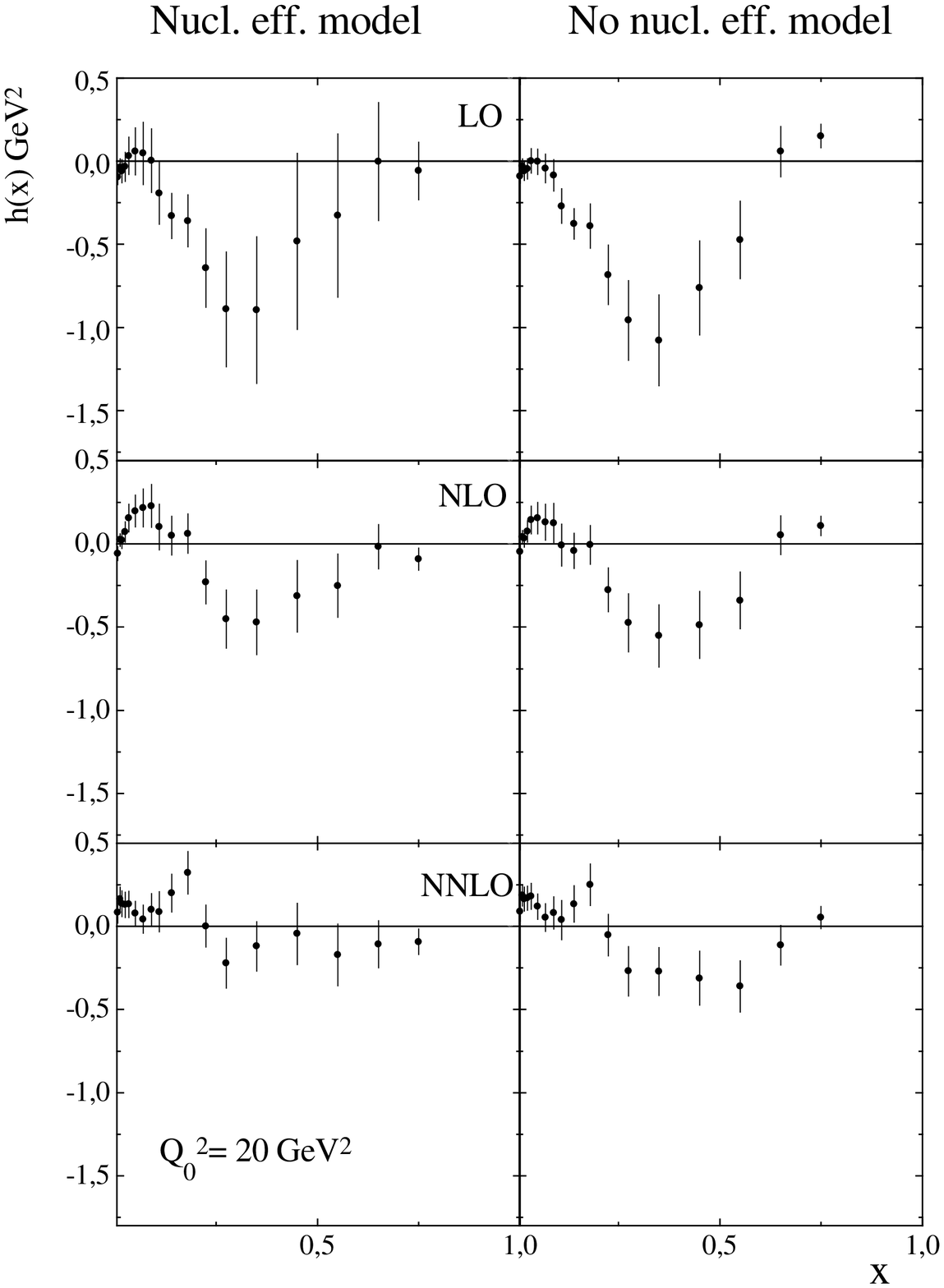}
\end{center}
\protect\caption{
The function $H_3(x)$ in the units of GeV$^2$, 
which describes the strength of the higher twist
term in the $xF_3$ structure function as extracted from the fit to the
CCFR neutrino data.
The labels on the figure indicate the level to which the perturbation theory
analysis of $xF_3^{LT}$ is done. Only statistical errors are shown.
\protect\label{1} }
\end{figure}

Our results are shown in Fig.\ref{1}. 
One can observe from Fig.\ref{1} that 
the scale which determines 
the magnitude of the HT term is 1\,GeV$^2$.
One also sees that the magnitude of the HT term
depends on the level to which
the perturbation theory analysis of $F_3^{LT}$ is done.
The more perturbative corrections are included into the evolution equation,
the less room is left for the function $H_3(x)$.
The separation of nuclear effects from  data leads to a further suppression of
the HT term $H_3(x)$. The effect of nuclear corrections is most pronounced 
at large $x$, where we
observe a systematic reduction of $H_3(x)$ as compared with no-nuclear-effects analysis.
Nuclear corrections result in the decrease of the values of the function $H_3(x)$ at
$x>0.6$. 
We also found that the separation of nuclear effects causes the $\Lambda_{\overline{MS}}$
to decrease for about 10\% leading to the shift of the value of 
$\alpha_S(M_Z)$ for about $2\cdot 10^{-3}$.
As a final remark, we comment that a reduction of the HT terms because of nuclear effects
at large $x$ has recently been observed in the analysis of charged lepton DIS data on
proton and deuteron targets \cite{akl}.%
\footnote{
However, the HT term in the structure function $F_2$ of \cite{a,akl}
appears to be an order of magnitude less than the central points of $H_3$ of \cite{kps,ks}. 
Furthermore, the effect of strong 
correlation between the order of pQCD analysis and the magnitude of the HT term,
first observed in \cite{kps} for neutrino data, is not
seen in the recent analysis of \cite{a}.
}

\section*{Acknowledgments}
The author is grateful to the organizers of the Moriond-2002 meeting for support.
This work was supported in part by the RFBR project no. 00-02-17432.


\section*{References}
\begin{thebibliography}{99}
\bibitem{a}
S.I. Alekhin, \Journal{\PRD}{63}{094022}{2001}; see also this Proceedings.
\bibitem{mv}
A. Milsztajn and M. Virchaux, \Journal{\PLB}{274}{221}{1992}.
\bibitem{kps}
A.L. Kataev, G. Parente, and A.V. Sidorov, \Journal{\NPB}{573}{405}{2000}.
\bibitem{akl}
S.I. Alekhin, S.A. Kulagin, and S. Liuti, to be published.

\bibitem{ks}S.A. Kulagin and A.V. Sidorov, \Journal{\EPJA}{9}{261}{2000}.



\bibitem{ccfr}
CCFR-NuTeV Collab., W.G. Seligman et al., \Journal{\PRL}{79}{1213}{1997}.

\bibitem{GL}
S.G. Gorishny and S.A. Larin, \Journal{\PLB}{172}{109}{1986}.

\bibitem{binding}
S. V. Akulinichev, S. A. Kulagin, and G. M. Vagradov, \Journal{\PLB}{158}{485}{1985};
S. A. Kulagin, \Journal{\NPA}{500}{653}{1989};
S. A. Kulagin, \Journal{\NPA}{640}{435}{1998}.

\bibitem{off-shell}
S.A. Kulagin, G. Piller and W. Weise, \Journal{\PRC}{50}{1154}{1994}.



\end{thebibliography}
\end{document}